\documentstyle[pra,aps,epsfig,eqsecnum,amssymb,amsmath,amsthm]{revtex}
\tighten
\newcounter{theo}
\newcounter{defi}
\newcounter{lemm}
\begin{document}
\clearpage
\preprint{}
\draft
\title{Some bounds for quantum copying}

\author{ A. E. Rastegin }
\address{Irkutsk State University, Irkutsk,
         664003, Russia \\
{\rm rast@api.isu.runnet.ru}}

\setcounter{page}{1}
\setcounter{theo}{0}
\setcounter{defi}{0}
\setcounter{lemm}{0}
\maketitle
\begin{abstract}
We propose new optimality criterion for the estimation of
state-dependent cloning. We call this measure  the relative error
because the one compares the errors in the copies with
contiguous size taking into account the similarity of states to be
copied. A copying transformation and dimension of state space are not
specified. Only the unitarity of quantum mechanical transformations
is used. The presented approach is based on the notion of the angle
between two states. Firstly, several useful statements simply
expressed in terms of angles are proved. Among them there are the
spherical triangle inequality and the inequality establishing the
upper bound on the modulus of difference between probability
distributions generated by two any states for an arbitrary
measurement. The tightest lower bound on the relative error is then
obtained. Hillery and Bu\v{z}ek originally examined an approximate
state-dependent copying and obtained the lower bound on the absolute
error. We consider relationship between the size of error and the
corresponding probability distributions and obtain the tightest lower
bound on the absolute error. Thus, the proposed approach
supplements and reinforces the results obtained by Hillery and
Bu\v{z}ek. Finally, the basic findings of investigation for the
relative error are discussed.
\end{abstract}


\pagenumbering{arabic}
\setcounter{page}{1}
\protect\section{Introduction}

One of the fundamental distinctions of the quantum world from
the classical world is expressed by the no-cloning theorem
\cite{wootters}. This result was generalized and extended in
paper \cite{barnum}. There are many applications of this statement,
for example, to the quantum cryptography. In some protocols Alice and
Bob encode the bits 1 and 0 into two non-orthogonal pure states
\cite{bennett}. Eve, a paradigmatical eavesdropper, then has
the following problem. Particle 1 (original mode) is secretly prepared
in some state from a set
${\mathfrak{A}}=\{|\phi\rangle,|\psi\rangle\}$ of two states.
Particle 2 (copy mode) is in some standard state $|0\rangle$.
Eve's action is to allow subsystem "1+2" to interact unitarily with
auxiliary system CM (copying machine) in some standard state
$|m\rangle\,$. Eve tries to implement a unitary process
\begin{equation*}
\forall \
|s\rangle \in {\mathfrak{A}} \ {\bf :} \quad {\rm U} \ |s\rangle
\otimes |0\rangle \otimes |m\rangle = |s\rangle \otimes
|s\rangle \otimes |k^{(s)}\rangle \!\ .
\end{equation*}
But the no-cloning theorem forbids such a "ideal copying" (except
when states $|\phi\rangle$ and $|\psi\rangle$ are orthogonal or
identical). It is clear, Eve will want to make the process output as
near to the ideal output as possible. How well Eve can do?

This question was originally considered by Hillery and Bu\v{z}ek in
\cite{hillery}. They examined approximate cloning machines destined
for copying of prescribed two non-orthogonal states. In paper
\cite{brass} such devices were called 'state-dependent cloners'.
Writers of \cite{brass} introduced the notion of 'global fidelity'
and constructed the optimal symmetric state-dependent cloner which
optimizes the global fidelity. Of course, the evaluation of
quality for one or another of potential cloners is dependent on the
used measure of 'closeness' to the ideality. As is readily seen from
the text, papers \cite{hillery} and \cite{brass} are based on the
distinct optimality criteria. Measure used by Hillery and Bu\v{z}ek
can be named 'absolute error' of copying of two-state set. In work
\cite{brass} the global fidelity was maximized.

However, a degree of similarity for states those are
subject to cloning is taken into account neither by the absolute
error nor by the global fidelity. It is small wonder, because these
measures are simply related. Meanwhile, the consideration of
similarity for states to be copied may have practical sense. In
general, it would be useful to have some set of optimality criteria
every of those elucidates one or another of facets of the
state-dependent cloning. In this work we propose
the new criterion for estimation of quality for the
state-dependent cloning. We call this criterion the relative
error because the one compares the errors in the copies with
contiguous size that is not independent of degree of similarity for
states to be copied. The lower bound on the relative error is
obtained. In our analysis we do not specify the dimension
of particle state space, although with applied viewpoint the
two-dimensional case is most interesting. We also
supplement and reinforce the results obtained by Hillery and
Bu\v{z}ek \cite{hillery}. In particular, the relationship between the
size of error and the absolute value of the deviation of the
resulting probability distribution from the desired probability
distribution is shown, as well as stronger bound is obtained.

In the examination all the state vectors are normalized to unity.
{\it In natura} non-unit vectors will sometimes occur, and this cases
will be expressly stated. As is customary, the norm of the vector
$|\Phi\rangle$ is defined as
$\,\|\:\!|\Phi\rangle\|=[\langle\Phi|\Phi\rangle]^{1/2}\,$.

\protect\section{Preliminary lemmas}

For comparing states $|\Phi\rangle$ and $|\Psi\rangle$
we shall use the angle between these vectors. Many relations can be
naturally expressed in terms of angles. In addition, the calculations
are simplified by the use of angles.

\newtheorem{d1}[defi]{Definition}
\begin{d1}
Angle $\delta(\Phi,\Psi)\in[0;\pi/2]$ between two
unit vectors $|\Phi\rangle$ and $|\Psi\rangle$ is defined as
\begin{equation*}
\delta(\Phi,\Psi)
\ {\mathop=^{{\rm def}}} \
\arccos\left(
\bigl| \langle\Phi | \Psi\rangle \bigr| \right) \!\ .
\end{equation*}
\end{d1}

This definition is naturally extended on the case of non-unit
vectors. It is obvious that function $\delta(\circ\,,\circ)$
is symmetric. For brevity we will also often write
$\delta_{\Phi\Psi}\,$. Expression $\delta_{\Phi\Psi}=0$ is
equivalent to the indentity of states $|\Phi\rangle$ and
$|\Psi\rangle$. Let us now prove two statements connected
with each other. These relations can be useful in various contexts.

\newtheorem{l1}[lemm]{Lemma}
\begin{l1}
For any triplet
$\{|\Phi\rangle,|\Upsilon\rangle,|\Psi\rangle\}$ of unit vectors,
\begin{equation}
\cos\delta_{\Phi\Psi}\leq
\cos(\delta_{\Phi\Upsilon}-\delta_{\Upsilon\Psi})
\!\ ,
\label{cosmin}
\end{equation}
with equality only if the triplet is coplanar.
\end{l1}

\begin{proof}
It is sufficient to consider the case, in which the
triplet has no identical states. Let vectors
$|\Theta\rangle$ and $|\Omega\rangle$ be such that
$\>|\Theta\rangle\perp|\Upsilon\rangle\>$ and
$\>|\Theta\rangle\in{\rm span}\{|\Phi\rangle,|\Upsilon\rangle\}\>$,
$\>|\Omega\rangle\perp|\Upsilon\rangle\>$ and
$\>|\Omega\rangle\in{\rm span}\{|\Psi\rangle,|\Upsilon\rangle\}\>$.
These vectors can be always constructed from non-collinear pairs
$\{|\Phi\rangle,|\Upsilon\rangle\}$ and
$\{|\Psi\rangle,|\Upsilon\rangle\}$ by the Gram--Schmidt
orthogonalization:
\begin{equation}
\begin{split}
 & |\Theta\rangle = \Bigl\{
|\Phi\rangle - |\Upsilon\rangle\langle\Upsilon | \Phi\rangle
\Bigr\} \Big/
\sqrt{ 1 - \bigl|\langle\Upsilon|\Phi\rangle\bigr|^2}
\!\ , \\
 & |\Omega\rangle = \Bigl\{
|\Psi\rangle - |\Upsilon\rangle\langle\Upsilon | \Psi\rangle
\Bigr\} \Big/
\sqrt{ 1 - \bigl|\langle\Upsilon|\Psi\rangle\bigr|^2}
\!\ .
\end{split}
\label{thome}
\end{equation}
Last equalities can be rewritten as
\begin{align*}
|\Phi\rangle & = |\Upsilon\rangle\langle\Upsilon | \Phi\rangle +
\sin\delta_{\Phi\Upsilon} |\Theta\rangle \!\ , \\
|\Psi\rangle & = |\Upsilon\rangle\langle\Upsilon | \Psi\rangle +
\sin\delta_{\Upsilon\Psi} |\Omega\rangle \!\ ,
\end{align*}
Applying the triangle inequality for complex numbers to equality
\begin{equation*}
\langle\Phi|\Psi\rangle=
\langle\Phi | \Upsilon\rangle\langle\Upsilon | \Psi\rangle
+\sin\delta_{\Phi\Upsilon}\sin\delta_{\Upsilon\Psi}
\langle\Theta|\Omega\rangle
\end{equation*}
and taking into account that in line with the Schwarz inequality
$|\langle\Theta|\Omega\rangle|\leq1$, we then get
\begin{align*}
\bigl|\langle\Phi|\Psi\rangle\bigr| &
\leq \cos\delta_{\Phi\Upsilon}\cos\delta_{\Upsilon\Psi} +
\sin\delta_{\Phi\Upsilon}\sin\delta_{\Upsilon\Psi}
\bigr|\langle\Theta|\Omega\rangle\bigl| \\
 & \leq \cos(\delta_{\Phi\Upsilon}-\delta_{\Upsilon\Psi})
\!\ .
\end{align*}
The maximal value
$\:\cos(\delta_{\Phi\Upsilon}-\delta_{\Upsilon\Psi})\:$ is reached
only if unit vectors $|\Theta\rangle$ and $|\Omega\rangle$ are
collinear. The latter together with (\ref{thome}) implies that there
is linear dependence for the triplet and the triplet is coplanar.
\end{proof}

\newtheorem{l2}[lemm]{Lemma}
\begin{l2}
For any triplet
$\{|\Phi\rangle,|\Upsilon\rangle,|\Psi\rangle\}$ of unit vectors,
\begin{equation}
\delta(\Phi,\Upsilon)\leq
\delta(\Phi,\Psi)+\delta(\Upsilon,\Psi)
\!\ ,
\label{cosplu}
\end{equation}
with equality only if the triplet is coplanar.
\end{l2}

\begin{proof}
{\it Ex adverso}, let
$\,\delta_{\Phi\Upsilon}>\delta_{\Phi\Psi}+\delta_{\Upsilon\Psi}\,$.
Taking into account the angle range of values, we get
$\:0\leq\delta_{\Phi\Psi}<\delta_{\Phi\Upsilon}-
\delta_{\Upsilon\Psi} \leq\pi/2\:$ and
$\:\cos\delta_{\Phi\Psi} >
\cos(\delta_{\Phi\Upsilon}-\delta_{\Upsilon\Psi})\:$.
But this contradicts to lemma 1, so that (\ref{cosplu}) is
true. Equality in (\ref{cosplu}) gives equality in (\ref{cosmin}),
that is possible only if the triplet is coplanar.
\end{proof}

The last statement is the base of our approach to obtaining
of bounds for state-dependent cloning. Inequality (\ref{cosplu}) can
be called 'spherical triangle inequality'. This becomes obvious if to
represent unit vectors $|\Phi\rangle$, $|\Upsilon\rangle$ and
$|\Psi\rangle$ by three points on the sphere with unit radius. Then
quantities $\delta(\Phi,\Upsilon)$, $\delta(\Phi,\Psi)$ and
$\delta(\Upsilon,\Psi)$ are the sides of the spherical triangle
formed by these points. Equality
$\:\delta_{\Phi\Upsilon}=\delta_{\Phi\Psi}+\delta_{\Upsilon\Psi}\:$
holds when the spherical triangle is degenerated into arc of
great circle and point $\Psi$ is found between points $\Phi$ and
$\Upsilon$. In this case triplet
$\{|\Phi\rangle,|\Upsilon\rangle,|\Psi\rangle\}$ is coplanar and
vector $|\Psi\rangle$ lies between $|\Phi\rangle$ and
$|\Upsilon\rangle$.

Below we shall show that if the angle between two states is small,
then the probability distributions generated by them for an
arbitrary measurement are close to each other. The measurement
over the system in state $|S\rangle$ produces result $R$ with
probability (p.e., see \cite{sudbery})
\begin{equation}
P(R\,|\,S) = \langle S| \,\Pi\, |S\rangle \!\ ,
\label{prob}
\end{equation}
where $\Pi$ is the operator of the orthogonal
projection onto the corresponding subspace.

\newtheorem{l3}[lemm]{Lemma}
\begin{l3}
For arbitrary triplet $\{|\Theta\rangle,|\Phi\rangle,|\Psi\rangle\}$
of unit vectors,
\begin{equation}
\left| \bigl| \langle \Theta |\Phi\rangle
\bigr|^2 - \bigl| \langle \Theta |\Psi\rangle \bigr|^2 \right| \leq
\sin\delta_{\Phi\Psi} \!\ .
\label{prob1}
\end{equation}
\end{l3}

\begin{proof}
Using lemma 1 and standard trigonometric formula (see \cite{handbook})
\begin{equation*}
\cos^2\!\alpha - \cos^2\!\beta =
-\sin(\alpha+\beta)\sin(\alpha-\beta) \!\ ,
\end{equation*}
we have
\begin{align*}
 & \bigl| \langle \Theta |\Phi\rangle \bigr|^2
- \bigl|\langle\Theta|\Psi\rangle\bigr|^2
\leq
\cos^2(\delta_{\Phi\Psi} - \delta_{\Psi\Theta}) -
\cos^2\!\delta_{\Psi\Theta}= \\
 & = \sin\delta_{\Phi\Psi}
\sin(2\delta_{\Psi\Theta} - \delta_{\Phi\Psi})
\leq \sin\delta_{\Phi\Psi}
\!\ .
\end{align*}
We further get by a parallel argument
\begin{equation*}
\bigl|\langle\Theta|\Psi\rangle\bigr|^2-
\bigl|\langle\Theta|\Phi\rangle\bigr|^2
\leq \sin\delta_{\Phi\Psi} \!\ ,
\end{equation*}
and the two last inequalities give (\ref{prob1}).
\end{proof}

\newtheorem{l4}[lemm]{Lemma}
\begin{l4}
For an arbitrary projector $\Pi$ and two any states
$|\Phi\rangle$ and $|\Psi\rangle$,
\begin{equation}
\bigl|
\langle\Phi| \,\Pi\, |\Phi\rangle -
\langle\Psi| \,\Pi\, |\Psi\rangle
\bigr| \leq \sin\delta_{\Phi\Psi}
\!\ .
\label{prob3}
\end{equation}
\end{l4}

\begin{proof}
It is sufficient to consider the case, in which vectors
$\Pi|\Phi\rangle$ and $\Pi|\Psi\rangle$ are non-zero
(other cases are reduced to lemma 3). Let us introduce two unit
vectors
$\:|\Theta\rangle=
\Pi|\Phi\rangle\,\big/\,\|\Pi|\Phi\rangle\|\:$ and
$\:|\Omega\rangle =
\Pi|\Psi\rangle\,\big/\,\|\Pi|\Psi\rangle\|\:$,
which lie in the subspace generated by $\Pi$.
We then have
\begin{equation}
\langle\Phi| \,\Pi\, |\Phi\rangle =
\bigl| \langle \Theta |\Phi\rangle
\bigr|^2 \!\ ,
\quad
\langle\Psi| \,\Pi\, |\Psi\rangle =
\bigl| \langle \Omega |\Psi\rangle
\bigr|^2
\!\ .
\label{pipip}
\end{equation}
We now express $|\Phi\rangle$ and $|\Psi\rangle$ as
$$
|\Phi\rangle=|\Theta\rangle\langle\Theta|\Phi\rangle
+|\Phi_{\perp}\rangle \!\ , \quad
|\Psi\rangle=|\Omega\rangle\langle\Omega|\Psi\rangle
+|\Psi_{\perp}\rangle \!\ ,
$$
where (generally, non-unit) vectors
$\;|\Phi_{\perp}\rangle\;$ and
$\;|\Psi_{\perp}\rangle\;$
are orthogonal to
${\rm span}\{|\Theta\rangle,|\Omega\rangle\}$.
Because
\begin{equation*}
\bigl|
\langle\Theta
|\Psi\rangle
\bigr|=
\bigl|
\langle\Theta
|\Omega\rangle
\bigr|\;
\bigl|
\langle\Omega|\Psi\rangle
\bigr|
\leq
\bigl|
\langle\Omega|\Psi\rangle
\bigr|
\!\ ,
\end{equation*}
we get by the use of lemma 3 the following inequality:
$$
\bigl| \langle \Theta |\Phi\rangle
\bigr|^2 -
\bigl| \langle \Omega |\Psi\rangle
\bigr|^2
\leq
\bigl| \langle \Theta |\Phi\rangle
\bigr|^2 - \bigl| \langle \Theta |\Psi\rangle \bigr|^2
\leq \sin\delta_{\Phi\Psi}
\!\ .
$$
We next have by a parallel argument
\begin{equation*}
\bigl| \langle \Omega |\Psi\rangle
\bigr|^2 -
\bigl| \langle \Theta |\Phi\rangle
\bigr|^2
\leq \sin\delta_{\Phi\Psi}
\!\ .
\end{equation*}
The two last inequalities and relations (\ref{pipip}) give
(\ref{prob3}) right away.
\end{proof}

Applying (\ref{prob}) and lemma 4, we then have the relation
\begin{equation}
\bigl|
P(R\,|\,\Phi) - P(R\,|\,\Psi)
\bigr| \leq \sin\delta_{\Phi\Psi}
\label{prob4}
\end{equation}
for an arbitrary measurement and two any states $|\Phi\rangle$ and
$|\Psi\rangle$. In accordance with (\ref{prob4}) small angle
between two states implies the closeness of probability distributions
generated by these states for any measurement. Thus, the angle
between two states gives a reasonable measure of closeness for two
pure states.

{\it Addendum} we shall apply (\ref{prob3}) to the quantum circuit
model, in which any unitary transformation is approximated by
a network formed from universal gates (see, for example,
\cite{ekert}). Let two unitary transformations U and V obey
\begin{equation}
\| {\rm U} - {\rm V} \| \leq \varepsilon \!\ ,
\label{netw}
\end{equation}
where the matrix norm is induced by the Euclidean norm of vectors.
Acting by ${\rm U}$ or ${\rm V}$ on some initial state
$|\Sigma\rangle$, we get final state
$|\Gamma\rangle={\rm U}|\Sigma\rangle$ or
$|\Upsilon\rangle={\rm V}|\Sigma
\rangle$ respectively. Relation
(\ref{netw}) means that
$\,\|\:\!|\Gamma\rangle-|\Upsilon\rangle\|\leq\varepsilon\,$,
whence
\begin{equation*}
2(1 - \cos\delta_{\Gamma\Upsilon})\leq
2\bigl(1 - {\rm Re}\langle \Gamma|\Upsilon\rangle\bigr)=
\| \:\! |\Gamma\rangle - |\Upsilon\rangle \|^2
\leq \varepsilon^2 \!\ .
\end{equation*}
Therefore,
$\:1-\cos^2\!\delta_{\Gamma\Upsilon}\leq\varepsilon^2(1-\varepsilon^2/4)\:$,
and for any measurement we then have
\begin{equation*}
\bigl|
P(R\,|\,\Gamma) - P(R\,|\,\Upsilon)
\bigr| \leq \varepsilon\> \sqrt{1 - \varepsilon^2/4}
\!\ .
\end{equation*}
Therefore, if ${\rm V}$ is substituted for ${\rm U}$ in a quantum
network, then the probability of arbitrary measurement outcome on the
final state is affected by at most $\varepsilon$.

\protect\section{Basic definitions}

Let ${\mathfrak{A}}=\{|\phi\rangle,|\psi\rangle\}$ be the set of two
pure states which we would like to copy. The action of the copying
machine can be expressed as
\begin{equation}
\forall \
|s\rangle \in {\mathfrak{A}} \ {\bf :} \quad
{\rm U}\ |s\rangle\otimes|0\rangle
\otimes |m\rangle =
|V^{(s)}\rangle \!\ ,
\label{abs}
\end{equation}
where $|V^{(s)}\rangle$ is in the state space of composite system
"1+2+CM". The unitarity of transformation ${\rm U}$ implies that
\begin{equation}
\langle\phi|\psi\rangle=
\langle V^{(\phi)}|V^{(\psi)}\rangle \!\ , \quad
\delta_{\phi\psi}=
\delta (V^{(\phi)},V^{(\psi)})
\!\ .
\label{abs1}
\end{equation}
In paper \cite{hillery} output $|V^{(s)}\rangle$ was
expressed as
\begin{equation}
\begin{split}
& |V^{(s)}\rangle=
|s\rangle\otimes|s\rangle\otimes|q^{(s)}\rangle +
|{\perp}^{(s)}\rangle
\!\ , \\
& |s\rangle\otimes|s\rangle\otimes|q^{(s)}\rangle
=\{|s\rangle\langle s|\otimes
|s\rangle\langle s|\otimes{\mathbf{1}}
\} |V^{(s)}\rangle \!\ ,
\end{split}
\label{buz}
\end{equation}
where $\mathbf{1}$ is the identity operator.
The unitarity of copying transformation imposes constraint
\begin{equation}
\|\>\! |q^{(s)}\rangle \|^2 +
\|\:\! |{\perp}^{(s)}\rangle \|^2 = 1
\!\ ,
\label{buz1}
\end{equation}
because the idempotency of projector
$\:
|s\rangle\langle s|\otimes|s\rangle\langle s|\otimes{\mathbf{1}}
\:$
gives
\begin{equation}
\{|s\rangle\langle s|\otimes
|s\rangle\langle s|\otimes{\mathbf{1}}
\} |{\perp}^{(s)}\rangle = 0
\label{perp}
\!\ .
\end{equation}
Hillery and Bu\v{z}ek introduced quantity
$\:X^{(s)}=\|\:\!|{\perp}^{(s)}\rangle\|\:$
as the size of error of state $|s\rangle$ copying. However,
the relationship between $X^{(s)}$ and the deviation of the
resulting probability distribution from the desired probability
distribution was not discussed in paper \cite{hillery}. We will below
study this relationship, and we will define in passing some importent
objects. Let us introduce magnitude
\begin{equation}
\delta^{(s)} = \inf \bigl\{ \
\delta(V^{(s)},s\otimes s\otimes k) \ \big| \
\langle k|k\rangle=1 \ \bigr\} \!\ .
\label{deltas}
\end{equation}
Using relation (\ref{perp}),
we see that the inner product of unit vectors $\langle V^{(s)}|$ and
$\>|s\rangle\otimes|s\rangle\otimes|k\rangle\>$ is equal to $\langle
q^{(s)}|k\rangle\>$. Because
$\,\|\>\!|k\rangle\|=1\,$, the Shwarz inequality gives
\begin{equation*}
\bigl| \langle q^{(s)}|k\rangle \bigr| \leq \|\>\!
|q^{(s)}\rangle \| \!\ ,
\end{equation*}
where the equality takes place if and only if
$\>|q^{(s)}\rangle=c\>|k\rangle\>$ for some complex number $\,c\,$.
The maximal value of the modulus of the inner product of two unit
vectors $\langle V^{(s)}|$ and
$\>|s\rangle\otimes|s\rangle\otimes|k\rangle\>$ corresponds to the
minimal value of angle $\>\delta(V^{(s)},s\otimes s\otimes k)\>$
between these vectors, so that if for some vector $|k\rangle$  infimum
(\ref{deltas}) is reached, then unit vector $|k\rangle$ and vector
$|q^{(s)}\rangle$ are collinear. For
$\,\|\>\!|q^{(s)}\rangle\|\not=0\,$ let us define vectors
\begin{equation}
|k^{(s)}\rangle
\ {\mathop=^{{\rm def}}} \
|q^{(s)}\rangle\bigm/\|\>\!|q^{(s)}\rangle\|
\ \ \ {\rm and} \ \ \
|{Id\,}^{(s)}\rangle
\ {\mathop=^{{\rm def}}} \
|s\rangle\otimes|s\rangle\otimes|k^{(s)}\rangle
\label{ideal}
\!\ .
\end{equation}
Infimum (\ref{deltas}) is reached for any vector
$\,|k\rangle=u\,|k^{(s)}\rangle\,$ with unit complex number $u\,$,
and $\delta^{(s)}$ is angle between unit vectors
$|V^{(s)}\rangle$ and $|{Id\,}^{(s)}\rangle$. Let Hermitian operator
${\rm A}$ describes some observable for particle 1. Its measurement
over particle in state $|s\rangle$ produces result $a$ with
probability
\begin{equation*}
p(a|s)=\langle s|\,\Pi_a|s\rangle \!\ ,
\end{equation*}
where $\Pi_a$ is the corresponding projector. If we shall now consider
this observable for composite system "1+2+CM", then the measurement of
such an observable over system "1+2+CM" in pure state $|V\rangle$
gives result $a$ with probability
\begin{equation*}
P(a\; {\rm for}\;
1\,|\,V)= \langle V|\:\Pi_a\!\otimes{\mathbf{1}}\otimes
{\mathbf{1}}\:
|V\rangle \!\ ,
\end{equation*}
where $\:\Pi_a\otimes{\mathbf{1}}\otimes{\mathbf{1}}\:$ is the
projector on the corresponding subspace of the composite system state
space. Such a representation for probability is equivalent to
expression ${\rm Tr}_1(\Pi_a\rho)$, where density operator $\rho$ of
particle 1 is the partial trace of operator
$\,|V\rangle\langle V|\,$
over subsystem "2+CM". For particle 2 we analogously have
\begin{equation*}
P(a\; {\rm for}\; 2\,|\,V)= \langle
V|\:{\mathbf{1}}\otimes\Pi_a\!\otimes{\mathbf{1}}\:  |V\rangle \!\ .
\end{equation*}
For state $|{Id\,}^{(s)}\rangle$ the probability of outcome
$a$ is
\begin{equation}
P(a\; {\rm for}\; j\,|\, {Id\,}^{(s)})=
\langle s|\,\Pi_a|s\rangle
=p(a|s) \!\ ,
\label{perfprob}
\end{equation}
where $j=1,2$ and $s=\phi,\psi$. Thus, $|{Id\,}^{(s)}\rangle$
corresponds to the ideal output. We can now relate angle
$\delta^{(s)}$ with objects introduced by Hillery and Bu\v{z}ek
\cite{hillery}. In line with definitions (\ref{deltas}) and
(\ref{ideal}) we have
\begin{equation*}
\cos\delta^{(s)}=
\bigl|\langle V^{(s)}|{Id\,}^{(s)}\rangle
\bigr|=\bigl|\langle q^{(s)}|k^{(s)}\rangle\bigr|=
\|\>\! |q^{(s)}\rangle\|
\!\ .
\end{equation*}
Then relation (\ref{buz1}) gives
$\>\|\:\!|{\perp}^{(s)}\rangle\|=\sin\delta^{(s)}\>$.
As inequality (\ref{prob4}), equality (\ref{perfprob}) and
definition of $X^{(s)}$ show,
\begin{equation}
\left| P(a\;
{\rm for}\; j\,|\, V^{(s)}) - p(a|s) \right| \leq X^{(s)} \!\ ,
\label{derpro}
\end{equation}
i.e. magnitude $X^{(s)}$ characterizes upon the whole the
deviation of the resulting probability distribution from the desired
probability distribution. Sum $\>X^{(\phi)}+X^{(\psi)}\>$
evaluates the total error of copying of set $\mathfrak{A}$.

\newtheorem{d2}[defi]{Definition}
\begin{d2}
Magnitude $\>AE({\mathfrak{A}})=X^{(\phi)}+X^{(\psi)}\>$ is the
absolute error of copying of set
${\mathfrak{A}}=\{|\phi\rangle,|\psi\rangle\}$.  \end{d2}

However, the absolute error does not take into account a degree of
similarity for states $|\phi\rangle$ and $|\psi\rangle$. Therefore,
we need a new measure for estimation of state-dependent cloning. The
notion of the relative error can be motived in the following way. Let
us take that we want distinguishing the input state by measurement
made on the copying output. If the modulus
$\,\bigl|\langle\phi|\psi\rangle\bigr|\,$ is close to 1 then the
lower bound on the absolute error is close to 0 and the upper bound
on the global fidelity is close to 1. In this case both criteria
assert that each copying output can be made near to which would be at
the ideality. We know both the ideal output
$|{Id\,}^{(\phi)}\rangle$ and the ideal output
$|{Id\,}^{(\psi)}\rangle$. At first sight
it seems that, comparing given output $|V^{(s)}\rangle$ to this ideal
output and to that one, we can recognize the input state not
uneasily. It would be a rashness to think so. Indeed, the closeness
for states $|\phi\rangle$ and $|\psi\rangle$ implies certain closeness
for the corresponding ideal outputs. But if so, are we able to decide
that given output should be related to this ideal output and not to
that one? To express this in quantitative form
we should use some measure of closeness for states
$|{Id\,}^{(\phi)}\rangle$ and $|{Id\,}^{(\psi)}\rangle$. Since
according to (\ref{prob4})
$$
\left| P(R\,|\, {Id\,}^{(\phi)}) -
P(R\,|\, {Id\,}^{(\psi)}) \right| \leq
\sin\delta({Id\,}^{(\phi)},{Id\,}^{(\psi)}) \!\ ,
$$
the quantity $\>\sin\delta({Id\,}^{(\phi)},{Id\,}^{(\psi)})\>$
provides such a measure. It stands to reason, this quantity is not
independent of similarity for states $|\phi\rangle$ and
$|\psi\rangle$. Are we willing to decide that, for example, given
output $|V^{(s)}\rangle$ should be related to ideal output
$|{Id\,}^{(\phi)}\rangle$ and not to $|{Id\,}^{(\psi)}\rangle$, when
$\>\sin\delta({Id\,}^{(\phi)},{Id\,}^{(\psi)})\>$ is small? The
closeness of $|V^{(\phi)}\rangle$ to $|{Id\,}^{(\phi)}\rangle$ and
the closeness of $|V^{(\psi)}\rangle$ to $|{Id\,}^{(\psi)}\rangle$
are described by the sizes
$X^{(\phi)}=\sin\delta^{(\phi)}$
and $X^{(\psi)}=\sin\delta^{(\psi)}$
respectively. It is not without significance that
$\>\sin\delta({Id\,}^{(\phi)},{Id\,}^{(\psi)})\>$ is size of the
same kind. Therefore, it is advisable to compare
the absolute error with pointed quantity. So, we have in a
reasonable way arrived at the definition of 'relative error'.

\newtheorem{d3}[defi]{Definition}
\begin{d3}
The ratio
$$
RE({\mathfrak{A}})=(X^{(\phi)}+X^{(\psi)})\,\big/\sin\delta({Id\,}^{(\phi)},{Id\,}^{(\psi)})
$$
is the relative error of copying of set
${\mathfrak{A}}=\{|\phi\rangle,|\psi\rangle\}$.
\end{d3}

We shall now derive the angle relations, from which bounds on the
errors are simply obtained. Using lemma 2 twice, we have
\begin{equation}
\delta({Id\,}^{(\phi)},{Id\,}^{(\psi)}) \leq
\delta^{(\phi)} + \delta^{(\psi)} +
\delta(V^{(\phi)},V^{(\psi)}) \!\ .
\label{condit}
\end{equation}
In accordance with the Schwarz inequality, there is
$$
\bigl|\langle {Id\,}^{(\phi)}|
{Id\,}^{(\psi)}\rangle
\bigr|=\bigl|\langle \phi|\psi\rangle \bigr|^2 \>
\bigl|\langle k^{(\phi)}|k^{(\psi)}\rangle \bigr|
\leq \bigl|\langle \phi|\psi\rangle \bigr|^2 \!\ ,
$$
whence we obtain $\:\delta({Id\,}^{(\phi)},{Id\,}^{(\psi)})
\geq\delta(\phi\otimes\phi,\psi\otimes\psi)\:$.
Therefore, $\:\delta(\phi\otimes\phi,\psi\otimes\psi)\leq
\delta^{(\phi)}+\delta^{(\psi)}+\delta(V^{(\phi)},V^{(\psi)})\:$,
or simply
\begin{equation}
\delta^{(\phi)} +
\delta^{(\psi)} \geq
\delta(\phi\otimes\phi,\psi\otimes\psi)
- \delta(\phi,\psi)
\label{keyunit}
\end{equation}
in line with Eq. (\ref{abs1}). Since
$\,0\leq\bigl|\langle\phi|\psi\rangle\bigr|\leq1\,$, there
is
$\,\delta(\phi\otimes\phi,\psi\otimes\psi)\geq\delta(\phi,\psi)\,$.
Eqs. (\ref{condit}) and (\ref{keyunit}) contain the restrictions
imposed by the laws of the quantum theory. In particular, the ones
allow to derive the lower bounds on both the relative error and
the absolute error.

\protect\section{Tightest lower bounds}

In this section we establish the lower bounds on both the relative
error and the absolute error. It should be pointed out that these
lower bounds are tightest. Indeed, we shall below describe a cloner
that reaches them. At first, let us formulate the result
that will be proved.

\newtheorem{t1}[theo]{Theorem}
\begin{t1}
The relative error $RE({\mathfrak{A}})$ of cloning for
set ${\mathfrak{A}}=\{|\phi\rangle,|\psi\rangle\}$ must be at least
as large as the quantity
\begin{equation} F(z) =z - z^2
\big/\sqrt{1+z^{2}} \!\ ,
\label{theore1}
\end{equation}
where $\>z=\bigl|\langle\phi|\psi\rangle\bigr|\>$.
\end{t1}

\begin{proof} In order to minimize $RE({\mathfrak{A}})$ the quantity
$\>\sin\delta({Id\,}^{(\phi)},{Id\,}^{(\psi)})\>$ must be as
increased as possible. We shall individually consider two cases, to
wit
(i) $\>\delta^{(\phi)}+\delta^{(\psi)}+
\delta_{\phi\psi}\leq\pi/2\>$ and
(ii) $\>\delta^{(\phi)}+\delta^{(\psi)}+
\delta_{\phi\psi}>\pi/2\>$.
Using Eqs. (\ref{abs1}) and (\ref{condit}), for the case (i) we have
$$
\sin\delta({Id\,}^{(\phi)},{Id\,}^{(\psi)})\leq
\sin(\delta^{(\phi)}+\delta^{(\psi)}
+\delta_{\phi\psi}) \!\ .
$$
The last relation, $\:\sin\delta^{(\phi)}+\sin\delta^{(\psi)}\geq
\sin(\delta^{(\phi)}+\delta^{(\psi)})\:$ and the trigonometric
formula for sine of difference give
\begin{equation}
RE({\mathfrak{A}})\geq
\cos\delta_{\phi\psi} -
\sin\delta_{\phi\psi}
\cot(\delta^{(\phi)}+\delta^{(\psi)}+\delta_{\phi\psi}) \!\ .
\label{cot}
\end{equation}
It must be stressed that the equality
\begin{equation}
\sin\delta^{(\phi)}+\sin\delta^{(\psi)}=\sin(\delta^{(\phi)}+\delta^{(\psi)})
\label{epsilbn}
\end{equation}
is necessary for the equality in Eq. (\ref{cot}). We want
minimizing the right-hand side of Eq. (\ref{cot}) in the interval
$\:\delta(\phi^{\otimes2},\psi^{\otimes2})\leq\delta^{(\phi)}+\delta^{(\psi)}+\delta_{\phi\psi}\leq\pi/2\:$
established by Eq. (\ref{keyunit}) and the case (i)
condition. Since the right-hand side of Eq. (\ref{cot}) increases as
the cotangent decreases and the cotangent is a decreasing function of
one's argument, the required minimum is reached at the left boundary
point of the interval stated above. Therefore, in the case (i)
\begin{equation}
RE({\mathfrak{A}})\geq
\sin\bigl(\delta(\phi^{\otimes2},\psi^{\otimes2}) -
\delta_{\phi\psi}\bigr)\,/
\sin\delta(\phi^{\otimes2},\psi^{\otimes2}) \!\ .  \label{icasmin}
\end{equation}
Moreover, we can see that in the case (i) the inequality
$\,AE({\mathfrak{A}})\geq\sin\bigl(\delta(\phi^{\otimes2},\psi^{\otimes2})
-\delta_{\phi\psi}\bigr)\,$ holds. In the case (ii) we have
$\>RE({\mathfrak{A}})\geq\sin\delta^{(\phi)}+\sin\delta^{(\psi)}\>$,
because $\>\sin\delta({Id\,}^{(\phi)},{Id\,}^{(\psi)})\leq1\>$
according to the definition of the angle. Next, the case (ii)
condition can be separated into two alternatives,
$\>\pi/2-\delta_{\phi\psi}<
\delta^{(\phi)}+\delta^{(\psi)}\leq\pi/2\>$
and $\>\pi/2<\delta^{(\phi)}+\delta^{(\psi)}\leq\pi\>$. The first
alternative contains
$$
RE({\mathfrak{A}})\geq\sin\bigl(\delta^{(\phi)}+\delta^{(\psi)}\bigr)
\geq\cos\delta_{\phi\psi} \!\ .
$$
In the second alternative the conditions
$\,\delta^{(s)}\leq\pi/2\,$ and
$\>\pi/2<\delta^{(\phi)}+\delta^{(\psi)}\leq\pi\>$ provide
$\>\sin\delta^{(\phi)}+\sin\delta^{(\psi)}\geq1\>$, that can be
easily verified by elementary methods. Thus, in the case (ii)
$\>RE({\mathfrak{A}})\geq\cos\delta_{\phi\psi}\>$ and
$\>AE({\mathfrak{A}})\geq\cos\delta_{\phi\psi}\>$. To sum up, we see that
the lower bound on the relative error is given by the right-hand side
of Eq. (\ref{icasmin}). Designating $\,z=\cos\delta_{\phi\psi}\,$,
hence $\,\cos\delta(\phi^{\otimes2},\psi^{\otimes2})=z^2\,$, the
right-hand side of (\ref{icasmin}) can be rewritten as $F(z)$.
\end{proof}

Thus, the relative error of copying of two-state set
$\mathfrak{A}$ must be at least as large as the right-hand side of
(\ref{theore1}) that is plotted in Fig.~1. We can see that in the
greater part of interval $z\in[0;1]$ the function increases and only
in the vicinity of the right boundary point the one becomes
decreasing. In general, the obtained result is clear. The more states
$|\phi\rangle$ and $|\psi\rangle$ are close to each other, the less
chances Eve has for the information extraction. Thus, the lower
bound on the relative error advises that Alice and Bob should make
encoding states $|\phi\rangle$ and $|\psi\rangle$ as close as
possible up to the vicinity in which the one slightly decreases.
However, the characteristics of a communication system can rather
limit a closeness for encoding states. In addition, if two
values $z_1$ and $z_2$ (where $z_1<z_2$) give the same lower bound on
the absolute error, value $z_2$ is more preffered, since for $z_2$
the lower bound on the relative error is larger than for $z_1\,$. As
it is shown, in this case the distinction between the bounds on the
relative error can be significant.

Next, the reasons used for proof of theorem 1 have
established the inequality
$\,AE({\mathfrak{A}})\geq
\sin\bigl(\delta(\phi^{\otimes2},\psi^{\otimes2})-\delta_{\phi\psi}\bigr)\,$
that can be rewritten in the following way.

{\bf Theorem 2}
{\it The absolute error of cloning for set
${\mathfrak{A}}=\{|\phi\rangle,|\psi\rangle\}$ has the lower bound:}
\begin{equation}
AE({\mathfrak{A}})
\geq z \sqrt{1-z^{4}} - z^2 \sqrt{1-z^{2}} \!\ .
\label{theore2}
\end{equation}

Thus, the absolute error of copying of two-state set $\mathfrak{A}$
must be at least as large as the right-hand side of (\ref{theore2}).
For small $z$ this function behaves as $z$, for small positive
$\,\xi=1-z\,$ one behaves as $(2-\sqrt{2})\sqrt{\xi}\,$. The maximal
value of the lower bound (\ref{theore1}) is equal to
$\,\sqrt{2/27}\approx0.272\,$ and occurs for
$\,z=1/\sqrt3\approx0.577\,$. The general bound
\begin{equation}
X^{(\phi)}+X^{(\psi)}
\geq 2 \left(\sqrt{1+z(1-z)} -1 \right)
\label{theoremhb}
\end{equation}
was obtained in paper \cite{hillery}.
The right-hand side of (\ref{theoremhb}) takes its
maximal value $\,\sqrt5-2\approx0.236\,$, when $z=1/2$. For $z=1/2$
our bound --- the right-hand side of (\ref{theore1}) --- is equal to
$\,\sqrt3\,(\sqrt5-1)/8\approx0.268\,$, and this value is not a
maximum. For small $z$ the right-hand side of (\ref{theoremhb})
behaves as $z$, for small positive $\,\xi=1-z\,$  one behaves as
$\xi$. Thus, we see that the lower bound given by (\ref{theore1}) is
stronger than the lower bound (\ref{theoremhb}). The distinction is
perceptable in the intermediate range of values of $z$ and for values
of $z$ close to 1. For example, at $z=4/5$ our bound is
approximatelly $1.5$ of the bound given by
(\ref{theoremhb}). The right-hand side of (\ref{theore1}) and the
right-hand side of (\ref{theoremhb}) are plotted as functions of
$z$ in Fig.~2 by the solid and dashed lines, respectively. As Fig.~2
shows, the bound (\ref{theoremhb}) is symmetric with respect to point
$z=1/2$, whereas the bound (\ref{theore1}) is asymmetric. Thus, the
presented approach has allowed to reinforce the lower bound derived
by Hillery and Bu\v{z}ek \cite{hillery}.

In principle, both lower bounds given by
theorems 1 and 2 can be reached without auxiliary device. Then a
unitary operator ${\rm U}$ acts on the Hilbert space of $2$ qubits:
\begin{align*}
 & |V^{(\phi)}\rangle={\rm U}\left\{
|\phi\rangle\otimes
|0\rangle\right\} \!\ ,\\
 & |V^{(\psi)}\rangle={\rm U}\left\{
|\psi\rangle\otimes
|0\rangle\right\} \!\ .
\end{align*}
The ideal output is $\,|{Id\,}^{(s)}\rangle=|s\otimes s\rangle\,$
for $s=\phi,\psi$, and there is
$\>\sin\delta({Id\,}^{(\phi)},{Id\,}^{(\psi)})=\sqrt{1-z^{4}}\>$.
It is clear that the necessary condition for minimization of the
errors is the equality in Eq. (\ref{keyunit}). According to lemma 2
the equality in Eq. (\ref{keyunit}) holds only if both final states
$|V^{(\phi)}\rangle$ and $|V^{(\psi)}\rangle$ lie in plane $\,{\rm
span}\{|\phi\otimes\phi\rangle,|\psi\otimes\psi\rangle\}\,$. This
is the necessary condition also for that the global fidelity is
maximazed \cite{brass}. Because unitary operations preserve
angles, we have
\begin{equation}
\delta(V^{(\phi)},V^{(\psi)})=
\delta(\psi\otimes0,
\psi\otimes0) \!\ . \label{vdnm}
\end{equation}
If states $|\phi\rangle$ and $|\phi\rangle$ are non-orthogonal and
non-identical then angle
$\,\delta(\phi^{\otimes 2},\psi^{\otimes 2})\,$
is larger than the right-hand side of Eq. (\ref{vdnm}) and
the ideal copying is impossible. In fact, it is impracticable that
angle between
$\:|\phi\otimes0\rangle\:$ and
$\:|\psi\otimes0\rangle\:$ should be properly
increased. To superpose the plane
$\:{\rm span}\{|\phi\otimes0\rangle,|\psi\otimes0\rangle\}\:$ onto
the plane $\,{\rm span}\{|\phi\otimes\phi\rangle,|\psi\otimes
\psi\rangle\}\,$ by rigid rotation ${\rm U}$ is at most that we can
achieve. The transformation with characteristics
\begin{align}
& {\rm span}\{|V^{(\phi)}\rangle,|V^{(\psi)}\rangle\}
={\rm span}\{|\phi\otimes\phi\rangle,|\psi\otimes\psi\rangle\}
\!\ , \label{symclo1} \\
& \delta^{(\phi)}=\delta^{(\psi)} =
\bigl(\delta(\phi^{\otimes2},\psi^{\otimes2})-\delta_{\phi\psi}\bigr)\big/2
\label{symclo2}
\end{align}
is the optimal symmetric state-dependent cloner constructed
in papers \cite{brass}. This
cloner produces equal errors for both states $|\phi\rangle$ and
$|\psi\rangle$. The absolute error $AE_{S}({\mathfrak{A}})$
is equal to the doubled sine of the angle stated in Eq.
(\ref{symclo2}). Using simple trigonometric formulae, we find that the
relative error for the cloner defined by Eqs. (\ref{symclo1}) and
(\ref{symclo2}) is equal to
$$
RE_{S}({\mathfrak{A}})=\sqrt{2}\,\left[
\:\frac{1+z+z^{2}}{1+z+z^{2}+z^{3}} -
\frac{1}{\sqrt{1+z^{2}}}
\:\right]^{1/2} \!\ .
$$
The defined by Eqs. (\ref{symclo1}) and (\ref{symclo2}) cloner does
not reach the equality in Eq. (\ref{epsilbn}) (except when states
$|\phi\rangle$ and $|\psi\rangle$ are orthogonal or identical, and,
hence, the angle stated in Eq. (\ref{symclo2}) is equal to zero).
Therefore, this symmetrical cloner minimizes neither the relative
error nor the absolute error. Note that for any symmetric
state-dependent cloner the relative error must be at least as large
as $RE_S({\mathfrak{A}})$.

We shall now propose an asymmetric cloner for which the relative
error of copying for set ${\mathfrak{A}}$ is rigorously equal to the
right-hand side of Eq. (\ref{theore1}). Such a optimal asymmetric
state-dependent cloner is defined by
\begin{align}
& {\rm span}\{|V^{(\phi)}\rangle,|V^{(\psi)}\rangle\}
={\rm span}\{|\phi^{\otimes 2}\rangle,|\psi^{\otimes 2}\rangle\}
\!\ , \label{asclo1} \\
& \delta^{(\phi)}=0 \wedge
\delta^{(\psi)} = \delta(\phi^{\otimes 2},\psi^{\otimes 2})
-\delta_{\phi\psi} \!\ .
\label{asclo2}
\end{align}
This cloner makes the ideal copying for one from prescribed pair
$\mathfrak{A}$ of states, i.e. the one is entirely asymmetric. It is
obvious that both the equality in Eq. (\ref{keyunit}) and the
equality in Eq. (\ref{epsilbn}) are reached. Therefore, for the
cloner defined by Eqs. (\ref{asclo1}) and (\ref{asclo2}) the relative
error $\,RE_A({\mathfrak{A}})=F(z)\,$ and the absolute error
$AE_A({\mathfrak{A}})$ is equal to the right-hand side of Eq.
(\ref{theore2}). In other words, the optimal asymmetric
state-dependent cloner minimizes both the relative and absolute
errors. So, without the symmetry requirement it is possible to build
the cloner with relative error that is smaller than relative error for
state-dependent cloner constructed in \cite{brass}.

Note that our asymmetric cloner, which makes the
ideal copying for one from prescribed pair of states, is not a special
example of the Wootters--Zurek cloner. The Wootters--Zurek
copying machine implements the ideal copying for the
orthogonal basis vectors \cite{buzek}. For pair
$\,{\mathfrak{A}}=\{|\phi\rangle,|\psi\rangle\}\,$ of non-orthogonal
states, we shall take the unit vector $|\omega\rangle$ such that
$\,|\omega\rangle={\rm span}\{|\phi\rangle,|\psi\rangle\}\,$ and
$\,\langle\phi|\omega\rangle=0\,$. Then $|\phi\rangle$ and
$|\omega\rangle$ are basis elements, and it may
be reasonable to consider the WZ--cloner such that
$\:|s\rangle\otimes|x\rangle\mapsto
|s\rangle\otimes|s\rangle\otimes|k^{(s)}\rangle\:$ for
$s=\phi,\omega$. Using simple calculations we find that
$\,RE_{W\!Z}({\mathfrak{A}})=\sqrt{3}\,z\big/\sqrt{1+z^2}\,$.
Therefore, the optimal asymmetric cloner defined by Eqs.
(\ref{asclo1}) and (\ref{asclo2}) differs from the Wootters--Zurek
copying machine.

\protect\section{Conclusion}

We have proposed the notion of the relative error which provides
new optimality criterion for the state-dependent cloning. We
have beforehand proved several useful statements those maintain our
approach. Among them there are the spherical triangle inequality and
the inequality establishing the upper bound on the modulus of
difference between probability distributions generated by two any
states for an arbitrary measurement. These relations can be useful in
various questions. Using physical reasons, the notion of the relative
error has been then introduced. The tightest lower bounds on the
absolute and relative errors of copying of the two-state set were
obtained. These bounds succeed the unitarity of quantum mechanical
transformations.

The lower bound on the relative error increases as function of $z$ in
the greater part of interval $z\in[0;1]$. Returning to cryptographic
example, it is possible to say roughly that the more states
$|\phi\rangle$ and $|\psi\rangle$ are close to each other the less
chances Eve has for the information extraction. It is not incurious
that in the vicinity of the right boundary point $z=1$ the lower
bound on the relative error becomes decreasing. Thus, this lower
bound advises that Alice and Bob should make encoding states
$|\phi\rangle$ and $|\psi\rangle$ as close as possible up to the
vicinity in which the one slightly decreases. However, the
characteristics of a communication system can rather limit a
closeness of encoding states.

As it is shown, the optimal symmetric state-dependent cloner, which
maximizes the global fidelity, reaches neither the lower bound on the
relative error nor the lower bound on the absolute error. We have
described the optimal asymmetric state-dependent cloner that
minimizes minimizes both the relative error and the absolute error.
It is worth noting that the global fidelity is optimized only if a
cloner is symmetric, whereas both the absolute and relative error are
optimized only if a cloner is entirely asymmetric.

It should be pointed out that the obtained results have only a
partial application to quantum communication problems because in the
reality a communication channel will inevitably suffer from noise
that will have caused the bits to evolve to mixed states.  Authors of
paper \cite{barnum} showed that noncommuting mixed states cannot be
ideally broadcast. It would be interesting to consider possible
limits on error in the case, where Eve's copying machine has as input
the original mode secretly prepared in one state from a set of two
mixed states.

Finally, we would like to point out that the above stated approach
can be applied to machines, which make multiple copies. Author
intends to examine bounds on errors for such a case in next paper.

\acknowledgments

I thank  Yuri V. Parfenov for helpful discussions and Anatoli A.
Jjenykh, Eugenia V.  Malchukova and Sergei I. Sinegovsky for help.

\newpage

\begin{figure}[t!] 
\vskip -20mm
\centering{\mbox{\epsfig{file=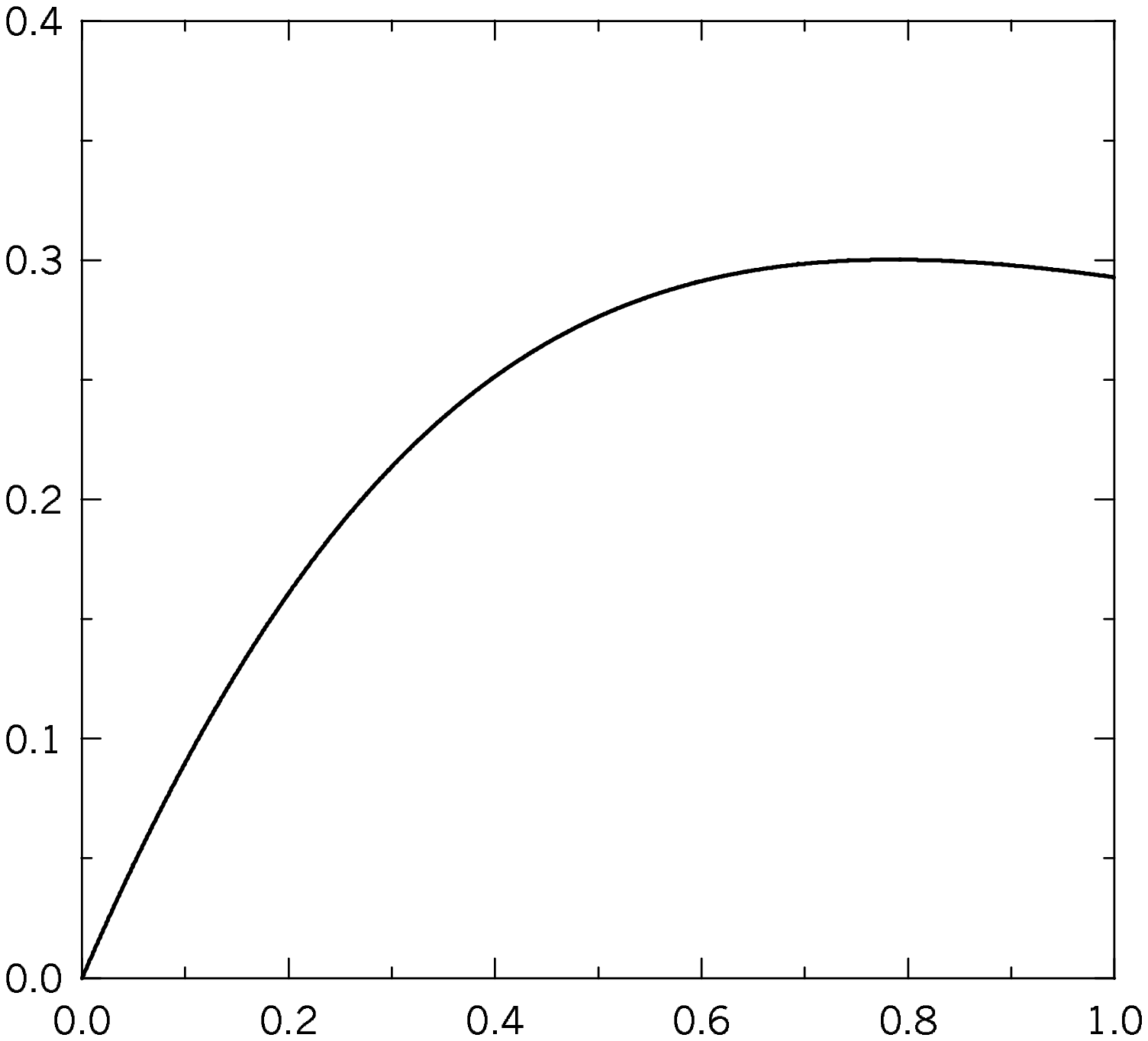,width=14.cm}}}
\vskip 5mm
\caption{The function $F(z)$ defined by the right-hand side of
(\ref{theore1}).}

\end{figure} 

\newpage

\begin{figure}[t!] 
\vskip -20mm
\centering{\mbox{\epsfig{file=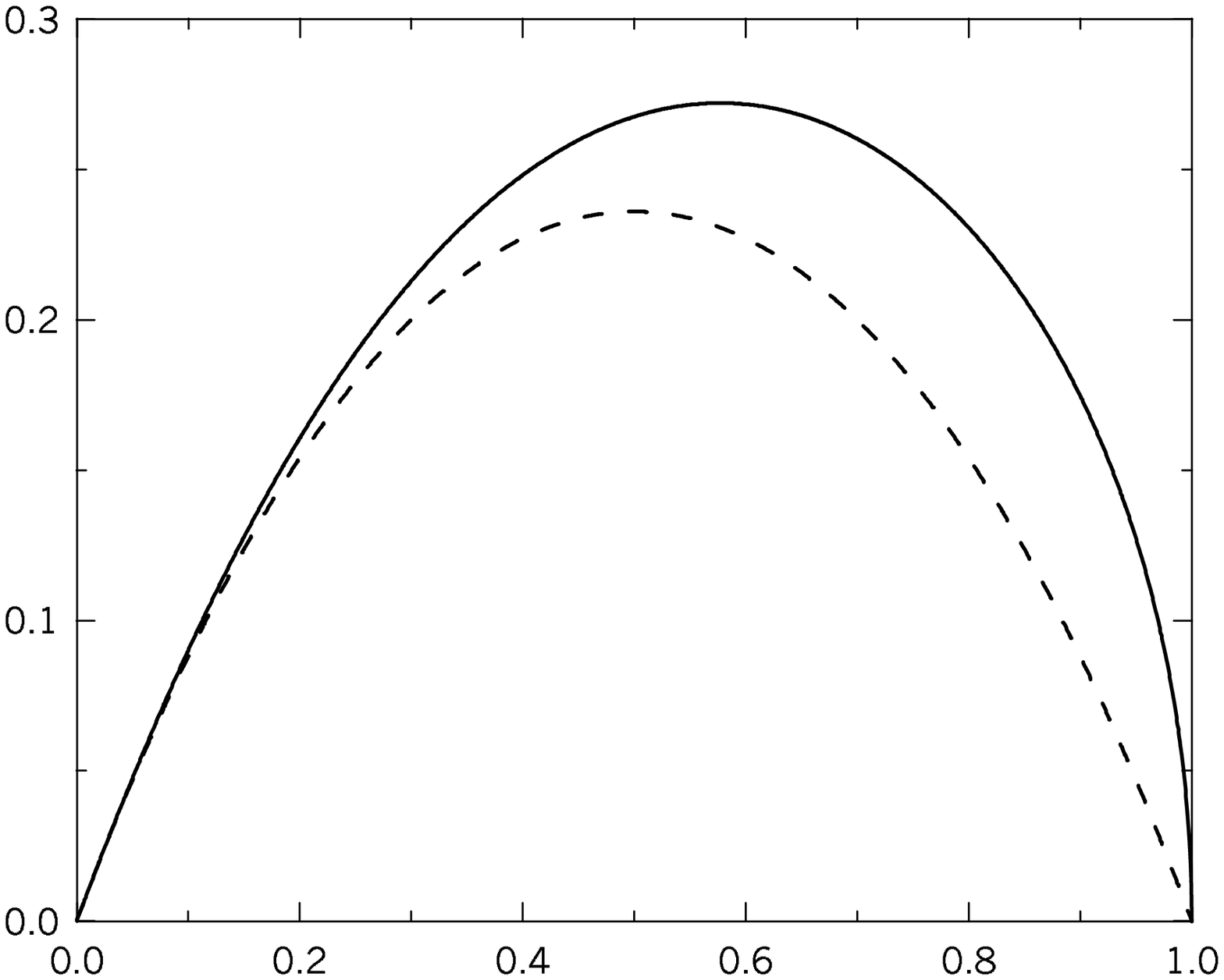,width=14.cm}}}
\vskip 5mm
\caption{The right-hand side of (\ref{theore2}) (solid line)
         and the right-hand side of (\ref{theoremhb}) (dashed line)
         as functions of $z$.}
\end{figure}

\end{document}